\documentclass[twocolumn,aps,prr,preprintnumbers,amsmath,amssymb,superscriptaddress,longbibliography]{revtex4-2}
\usepackage[utf8]{inputenc}
\usepackage{bm}% bold math
\usepackage{siunitx}
\usepackage{graphicx}% Include figure files
\usepackage{dcolumn}% Align table columns on decimal point
\usepackage[mathcal]{euscript}
\usepackage{color}
\usepackage[dvipsnames]{xcolor}
\usepackage{braket}
\def\beq{\begin{equation}}
\def\eeq{\end{equation}}
\newcommand{\redtransition}{$^{1\hspace{-0.3ex}}S_0$\,-\,$^{3\hspace{-0.3ex}}P_1\,$}
\newcommand{\bluetransition}{$^{1\hspace{-0.3ex}}S_0$\,-\,$^{1\hspace{-0.3ex}}P_1\,$}

\usepackage[english]{babel}
\usepackage{CJKutf8}
\usepackage{graphicx}
\usepackage{siunitx}
\usepackage[colorlinks=true,linkcolor=blue,citecolor=blue,urlcolor=blue,filecolor=blue]{hyperref}
\usepackage{tikz}
\usepackage{xcolor}
\usepackage{float}

\definecolor{lime}{HTML}{A6CE39}
\DeclareRobustCommand{\orcidicon}{
	\begin{tikzpicture}
	\draw[lime, fill=lime] (0,0) 
	circle [radius=0.16] 
	node[white] {{\fontfamily{qag}\selectfont \tiny ID}};
	\draw[white, fill=white] (-0.0625,0.095) 
	circle [radius=0.007];
	\end{tikzpicture}
	\hspace{-0.3mm}
}

\foreach \x in {A, ..., Z}{\expandafter\xdef\csname orcid\x\endcsname{\noexpand\href{https://orcid.org/\csname orcidauthor\x\endcsname}
			{\noexpand\orcidicon}}
}

\begin{document}

%\preprint{}

\title{Narrow-line imaging of single strontium atoms in shallow optical tweezers}

\author{Alexander Urech$\orcidA{}$}
\affiliation{Van der Waals-Zeeman Institute, Institute of Physics, University of Amsterdam, Science Park 904, 1098XH Amsterdam, The Netherlands}
\affiliation{QuSoft, Science Park 123, 1098XG Amsterdam, The Netherlands}
\author{Ivo H. A. Knottnerus$\orcidB{}$}
\affiliation{Van der Waals-Zeeman Institute, Institute of Physics, University of Amsterdam, Science Park 904, 1098XH Amsterdam, The Netherlands}
\affiliation{QuSoft, Science Park 123, 1098XG Amsterdam, The Netherlands}
\affiliation{Eindhoven University of Technology, P.O. Box 513, 5600MB Eindhoven, The Netherlands}%Center for Quantum Materials and Technology, Eindhoven University of Technology, P.O. Box 513, 5600MB Eindhoven, The Netherlands}
\author{Robert J. C. Spreeuw$\orcidC{}$}
\affiliation{Van der Waals-Zeeman Institute, Institute of Physics, University of Amsterdam, Science Park 904, 1098XH Amsterdam, The Netherlands}
\affiliation{QuSoft, Science Park 123, 1098XG Amsterdam, The Netherlands}
\author{Florian Schreck$\orcidD{}$}
\email[]{NarrowLineSrTweezerImaging@strontiumBEC.com}
\affiliation{Van der Waals-Zeeman Institute, Institute of Physics, University of Amsterdam, Science Park 904, 1098XH Amsterdam, The Netherlands}
\affiliation{QuSoft, Science Park 123, 1098XG Amsterdam, The Netherlands}

\date{\today}% It is always \today, today,

\begin{abstract} 

Single strontium atoms held in optical tweezers have so far only been imaged using the broad \bluetransition transition. 
For Yb, use of the narrow  (183\,kHz-wide) \redtransition transition for simultaneous imaging and cooling has been demonstrated in tweezers with a magic wavelength for the imaging transition. 
We demonstrate high-fidelity imaging of single Sr atoms using its even narrower (7.4\,kHz-wide) \redtransition transition. The atoms are trapped in \textit{non}-magic-wavelength tweezers. We detect the photons scattered during Sisyphus cooling, thus keeping the atoms near the motional ground state of the tweezer throughout imaging. 
The fidelity of detection is 0.9991(4) with a survival probability of 0.97(2). 
An atom in a tweezer can be held under imaging conditions for 79(3) seconds allowing for hundreds of images to be taken,  limited mainly by background gas collisions.
We detect atoms in an arrary of 36 tweezers with 813.4-nm light and trap depths of 135(20)\,$\mu$K. This trap depth is three times shallower than typically used for imaging on the broad \bluetransition transition. Narrow-line imaging opens the possibility to even further reduce this trap depth, as long as all trap frequencies are kept larger than the imaging transition linewidth.
Imaging using a narrow-linewidth transition in a non-magic-wavelength tweezer also allows for selective imaging of a given tweezer. As a demonstration, we selectively image (hide) a single tweezer from the array. This provides a useful tool for quantum error correction protocols.

\end{abstract}

%\pacs{}

%\keywords{}

\maketitle

\section{Introduction}
\label{sec:intro}%%%%%%%%%%%%%%%%%%%%%%%%%%%%%%%%%%%%%%%%%%%%%%%%%%%%%%%%%%%%%%
Optical tweezers have emerged as a powerful tool  for quantum applications. They enable state of the art quantum simulation and computation  \cite{Scholl2021_quantsim_browaeys,Ebadi2021_quantsim_lukin,Choi2021_quantSim_endres,Schine2021_Lattice}, high fidelity and long coherence time qubits  \cite{atom_computing_2021,Madjarov2020_Rydberg,Ma2021_Yb171_redImaging,Martin2021_CsRydbergGate},  quantum metrology  \cite{Norcia_Clock_2019,Madjarov2019_clock,Young2020_halfminuteClock}, quantum chemistry  \cite{Cairncross2021_molecule_rovibratoinal,Burchesky2021_polar_molecule}, among numerous other applications. 
Optical tweezers with alkaline-earth(-like) atoms, in particular with strontium and ytterbium, have been recently realized, offering new possibilities in expanding these applications \cite{cooper_alkaline-earth_2018,norcia_microscopic_2018,saskin_narrow-line_2019}.\

In all strontium tweezer experiments demonstrated so far, the fluorescence of single atoms on the broad (30\,MHz)\,\bluetransition transition at 461\,nm was recorded, while simultaneously cooling the atoms on the narrow (7.4\,kHz)\,\redtransition transition at 689\,nm \cite{cooper_alkaline-earth_2018,norcia_microscopic_2018,covey_2000-times_2019,Madjarov2019_clock,Madjarov2020_Rydberg,atom_computing_2021,Young2020_halfminuteClock,Schine2021_Lattice,Norcia_Clock_2019}. 
This `blue imaging' method allows for high-fidelity detection of single atoms in tweezers with high survival probability \cite{cooper_alkaline-earth_2018,norcia_microscopic_2018,covey_2000-times_2019}. 
However, blue imaging requires repumpers to close the 5s4d $^{1\hspace{-0.3ex}}D_2$ decay channel, which can only be done at tweezer wavelengths where also the $^{1\hspace{-0.3ex}}D_2$ state is trapped \cite{covey_2000-times_2019,Madjarov2020_Rydberg,Norcia_Clock_2019,atom_computing_2021}.\ 
Furthermore, the slightly higher scattering rate obtained in the blue imaging process ($\thicksim$\,75\,kHz) can only be used as long as the tweezers are sufficiently deep.
Any excess heating from the imaging process can then be cooled away after the image.
As the trap depth of the tweezer is reduced, the advantage of fast imaging is lost because the scattering rate must also be reduced to balance heating from imaging and cooling.
Reducing the tweezer trap depth has the advantages of decreased laser power requirement per tweezer (allowing for more tweezers using a given laser source) and  increasing metastable state lifetimes (reduced off-resonant scattering of tweezer light). \

A simpler method for imaging alkaline-earth(-like) atoms in tweezers is to use the narrow \redtransition transition for both cooling and imaging. 
Single atom detection by fluorescence imaging on a (less) narrow transition has previously been demonstrated in ytterbium for two different isotopes. In both cases tweezers with a magic-wavelength for the imaging transition were used \cite{saskin_narrow-line_2019, Ma2021_Yb171_redImaging}. \

Here we detect single $^{88}$Sr atoms using only the \redtransition\, transition for simultaneous`red imaging' and cooling.
We use optical tweezers that are \textit{non}-magic for the imaging transition, but magic for the Sr clock transition ($^{1\hspace{-0.3ex}}S_0$\,-\,$^{3\hspace{-0.3ex}}P_0\,$).
We detect the photons scattered during an \textit{attractive} Sisyphus cooling process \cite{covey_2000-times_2019}, thus  keeping the atoms near the motional ground state of the tweezer throughout imaging.\ 

Attractive Sisyphus cooling is possible at tweezer wavelengths where the excited state experiences a deeper trap depth than the ground state. 
More specifically, at our tweezer wavelength of 813.4\,nm, the excited state 
($\ket{e}\equiv\,^{3\hspace{-0.3ex}}P_1\,(|m_j|=1)$) confinement is 1.24 times greater than the ground state ($\ket{g}\equiv\,^1S_0$), see Fig.\,\ref{fig:PicsandExpSetup}(a). 
This cooling process can be very efficient with a proper choice of parameters, leading to a large reduction in energy per scattered photon, and a small number of scattered photons needed to cool the atom  \cite{ivanov_laser-driven_2011}.
In addition, due to the anharmonicity of the tweezer potential, atomic motion in all directions is coupled, allowing for a single radial cooling beam to remove energy from all directions  \cite{ivanov_laser-driven_2011,covey_2000-times_2019}.\ 

With balanced heating and cooling from the imaging process, the trap depth  can be significantly decreased, reducing the power required per tweezer. 
Additionally, red imaging can be performed without repumpers, since optical pumping to metastable states is much reduced. 
The only remaining pumping is due to off-resonant scattering of 813-nm tweezer light when the atom is in $\ket{e}$, which has a low rate that is even further decreased by using shallow tweezers.\ 

Imaging in shallow tweezers does limit the scattering rate that can be achieved without unacceptable atom loss, for both red and blue imaging. 
Shallow tweezer imaging therefore requires a longer exposure time for high-fidelity single atom detection compared to imaging in deeper tweezers.
We show for red imaging that the maximum scattering rate ($\Gamma/2\approx\,23$ kHz  for the \redtransition\ transition, where $\Gamma$= 2$\pi\times$7.4\,kHz is the transition linewidth) can be closely approached with proper modulation of the frequency and intensity of the cooling beam, while maintaining a lower temperature than when using blue imaging. 
This allows for a reduction of the trap depth by a factor of $\thicksim3$, while only marginally increasing the imaging duration and maintaining a near unity detection fidelity and survival probability.\ 

We proceed by presenting an overview of the experimental setup and the procedure for preparing single atoms in Sec.\,\ref{sec:overview}. 
In Sec\,\ref{sec:imaging}, we present our imaging method along with a description of the attractive Sisyphus cooling process, optimized  parameters, detection fidelity, and survival probabilities.  
In Section \ref{sec:siteselective} we demonstrate the ability to selectively image (dark out) a specific tweezer from the array and we conclude in
Sec.\,\ref{sec:conclusion}.

\section{Overview of experimental setup and procedure}
\label{sec:overview}
Similar to previously demonstrated strontium tweezer experiments, we load the optical tweezers with a small and random number of atoms from a magneto-optical trap (MOT) operating on the narrow  \redtransition\ transition  \cite{cooper_alkaline-earth_2018,norcia_microscopic_2018,covey_2000-times_2019,Madjarov2019_clock,Madjarov2020_Rydberg,atom_computing_2021,Young2020_halfminuteClock,Schine2021_Lattice,Norcia_Clock_2019}.  Our procedure for creating the MOT is similar to the one of \cite{stellmer_PreparationBEC}, but uses a reduced number of MOT beams to make space for a microscope objective and dynamically moves the MOT from a loading position into the objective focus, see Appendix A.\

We create two dimensional arrays of optical tweezers using a  phase-only spatial light modulator (SLM) to imprint a  phase onto an 813.4-nm laser beam creating an array of foci \cite{Scholl2021_quantsim_browaeys,Ebadi2021_quantsim_lukin}.
This array is then imaged onto the narrow linewidth MOT through an NA=0.5 microscope objective. 
An additional dynamically movable tweezer is created using the same microscope objective and a pair of crossed acousto-optic deflectors (AODs), see Appendix B.\

The tweezer trap depth used throughout this paper is 135(20)\,$\mu$K unless otherwise specified. For our $1/e^2$ tweezer waist of $\thicksim0.84\,\mu$m, the ground state radial (axial) trap frequencies are $\omega_{\rm radial}=\,43(3)$\,kHz ($\omega_{\rm axial}=\,6.6(5)$\,kHz), respectively. This trap depth is chosen such that the excited state axial trap frequency (7.3(6)\,kHz) is comparable to the linewidth of the \redtransition transition.

\begin{figure}
    \includegraphics[width=\columnwidth]{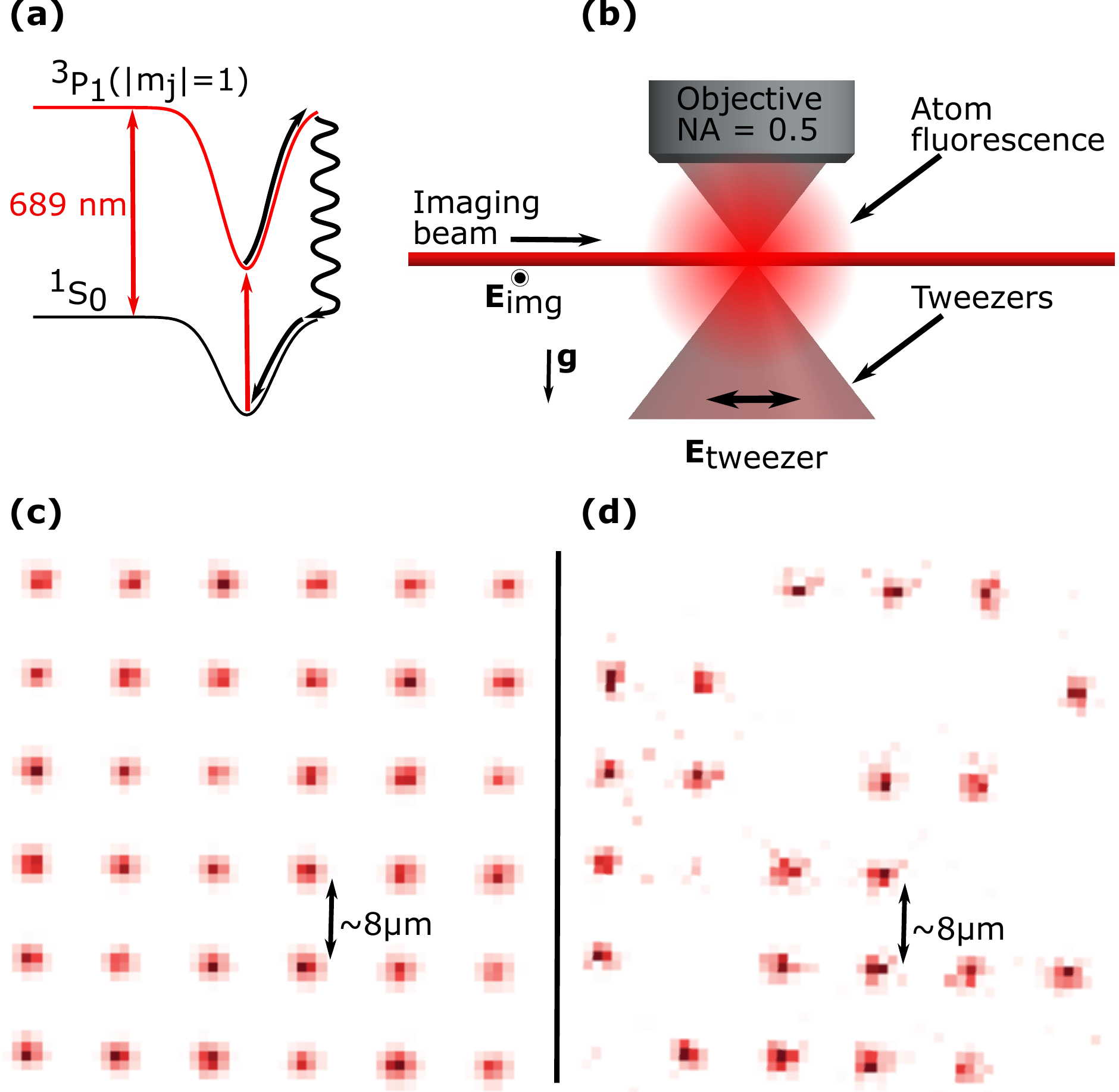}
    \caption{\textbf{(a)} A sketch of the Sisyphus cooling process at the root of our imaging technique. 
    The tweezer potential is deeper for the excited state ($^{3\hspace{-0.3ex}}P_1$) than for the ground state ($^1S_0$). The atom is preferentially excited near the bottom of the potential, then rolls up the steeper excited state potential before decaying.
    This leads to a reduction in energy per scattered photon related to the trap depth mismatch \cite{covey_2000-times_2019,ivanov_laser-driven_2011}
    \textbf{(b)} A simplified schematic of the experimental setup. 
    A high numerical aperture objective (NA = 0.5)  creates the optical tweezers and collects the atomic fluorescence.  
    A single beam (imaging beam) is  used for light assisted collisions, cooling, and imaging in the tweezer array. The polarization of both this imaging beam ($\textbf{E}_{\text{img}}$) and the SLM tweezer pattern ($\textbf{E}_{\text{tweezer}}$) are shown. 
    The direction of gravity with respect to the objective is also shown (\textbf{g}). 
    \textbf{(c)} Averaged fluorescence of strontium atoms in the 6$\times$6 array of tweezers used throughout the majority of this work. We collect photons scattered from the \redtransition $(|m_j|=1)$ transition during the cooling process in order to image the atoms in the array. 
    The image is the average of 100 experimental realizations using 500\,ms of exposure each.
    \textbf{(d)} Image obtained by one such experimental run. Approximately half of the tweezer traps are filled on average.}
    \label{fig:PicsandExpSetup} 
\end{figure}
Once the tweezers have been loaded and the MOT is switched off, a single (non-retro-reflected) 689-nm beam is used to address the tweezer array during all further experimental stages that need 689-nm light (light-assisted collisions, imaging, cooling, spectroscopy), see Fig.\,\ref{fig:PicsandExpSetup}(b). 
This beam, here simply called imaging beam, is linearly polarized perpendicular to the tweezer propagation axis to maximize the fluorescence into the microscope. 
Additionally, we tune the linear polarization of the tweezer light to match the propagation axis of the imaging beam. This maximises the $\sigma^\pm$ component of the imaging beam because we operate the tweezers at a 0\,G magnetic field making the tweezer polarization the dominant quantization axis.\

To prepare tweezers containing either a single or no atom, we use light-assisted collisions to induce pairwise loss, leaving either zero or one atom remaining in each tweezer \cite{Schlosser2002_CollisionalBlockade}. 
The imaging  light used in this process is tuned to a frequency between the Stark shifted resonance of the \redtransition $(|m_j|=1)$  transition and an electronically excited molecular state that is further red detuned and that asymptotically corresponds to the \ $^{3\hspace{-0.3ex}}P_1$ state  \cite{cooper_alkaline-earth_2018,zelevinsky_narrow_2006}. \

We perform imaging by collecting the scattered photons from the Sisyphus cooling process as presented in Sec.\,\ref{sec:imaging}.
The fluorescence is collected via the same microscope objective used to generate the tweezers, and then separated using a long pass dichroic mirror with 750-nm cutoff. The collected fluorescence light is sent onto an EMCCD camera  \cite{Hirsch2013_EMCCD,Bergschneider2018_Li_EMCCD}. 
The number of photons in 5$\times$5 pixel regions of interest (ROIs) around each tweezer center is summed. We collect photons for 100\,ms  in order to separate the single atom signal from the background noise of the camera.
This procedure leads to a histogram with two peaks, corresponding to zero and one atom in a tweezer, as shown in Fig.\,\ref{fig:HistoplusLifetime}(a). An atom is assumed to be in a tweezer if the photon number lies above a threshold located between the two peeks, see Sec.\,\ref{sec:imaging}(C).\

In the previously demonstrated blue imaging technique, the metastable $^{3\hspace{-0.3ex}}P_0$ and $^{3\hspace{-0.3ex}}P_2$ states must be repumped to the ground state during imaging because of decay of $^{1\hspace{-0.3ex}}P_1$ to those metastable states \cite{covey_2000-times_2019,Norcia_Clock_2019}. 
During red imaging, these repumpers are only used to compensate optical pumping into $^{3\hspace{-0.3ex}}P_{0,2}$ by the tweezer light that can happen when the atom is in the $^{3\hspace{-0.3ex}}P_1\,(|m_j|=1)$ state.
However, we find that this is unnecessary for the shallow traps used in this work. 
Nonetheless, we have the ability to repump the metastable states via the $^{3\hspace{-0.3ex}}S_1$ state using two lasers at 679 nm and 707 nm for the $^{3\hspace{-0.3ex}}P_0$ and $^{3\hspace{-0.3ex}}P_2$ states respectively.\ 
 
For all the results presented in this paper, we begin an experimental run by preparing single atoms using the above method followed by an initial image to determine which tweezers are filled. 
After this,we perform measurements as required by the experiment under consideration.
The average initial image of 100 preparations  for a 6$\times$6 array and an image of a single run is shown in Fig.\,\ref{fig:PicsandExpSetup}(c) and (d) respectively.\ 

All plots presented are for 100 repetitions of each experiment unless stated otherwise and the data is the average of all 36 tweezer sites. Taking into account the typical tweezer loading efficiency of $50\%$,  each data point consists of approximately 1800 realizations. The error bars for the entire paper show the standard deviation over the array and are dominated by variation originating from tweezer depth inhomogeneities across the array.\

%%%%%%%%%%%%%%%%%%%%%%%%%%%%%%%%%%%%%%%%%%%%%%%%%%%%%%%%%%%%%%%%
\section{Imaging via Sisyphus cooling}
\label{sec:imaging}
Our imaging and cooling  relies on the attractive Sisyphus cooling technique first proposed in  \cite{taieb_cooling_1994,ivanov_laser-driven_2011} and more recently observed experimentally in tweezer arrays  \cite{covey_2000-times_2019,Norcia_Clock_2019,atom_computing_2021} as well as in a continuous beam decelerator  \cite{chen_sisyphus_2019}.  
We can keep the scattering rate near maximum with near zero trap loss or heating in tweezers as shallow as 135(20)\,$\mu$K by intentionally keeping the imaged atoms slightly hotter than the coldest possible temperature, and by proper choice of imaging/cooling parameters.\ 

This section is ordered as follows. The required criterion for attractive Sisyphus cooling will be outlined in Sec.\,\ref{sec:imaging}(A). Sec.\,\ref{sec:imaging}(B) will present our characterization of the optimal cooling parameters. We will present our imaging method in Sec.\,\ref{sec:imaging}(C). We will end with the analysis of the detection fidelity and survival probability of our imaging process in Sec.\,\ref{sec:imaging}(D).

\subsection{Sisyphus cooling criteria}
\label{subsec:critera}
Attractive Sisyphus cooling relies on a trap depth mismatch between the excited and ground state  potentials as shown in Fig.\,\ref{fig:PicsandExpSetup}(a). In addition, three conditions must be fulfilled for the cooling to work.  First, the excited state of the atom must experience  stronger confinement than the ground state. 
Second, one must have the ability to excite the atom selectively from the bottom of the potential, and third the excited atom must have sufficient time to move away from the center of the potential before decaying  \cite{ivanov_laser-driven_2011}. 
The first condition is fulfilled in our setup by properly choosing the trapping wavelength, while the second and third  conditions can be fulfilled by using the narrow linewidth  \redtransition transition in strontium. \

The first condition is needed for the atoms to  lose kinetic energy by rolling up the steeper potential of the excited state before decaying.
This allows for a  reduction in potential energy on the order of the differential trap depth per scattering event. 
The narrow linewidth of the transition allows for the atoms to be selectively excited from the bottom of the trap if the differential trap depth is larger than the linewidth. 
The lifetime of the narrow transition is long enough to satisfy the third condition if the trap frequencies are larger than the linewidth. 
The atom will then  more likely  decay near the motional turning point, away from the center of the trap.\

\subsection{Optimal cooling}
To investigate the performance of our cooling/imaging technique  and find  optimal cooling parameters, we measure the temperature by the release and recapture method \cite{Tuchendler2008_release_recapture}.
We switch the tweezers off, wait a time $t_{\rm release}$ before turning them back on again, and then image the atoms to determine their survival fraction. Atoms are lost quicker when they are hotter. The temperature is determined by comparing the survival fraction for several values of $t_{\rm release}$ with Monte-Carlo atom trajectory simulations \cite{Tuchendler2008_release_recapture,covey_2000-times_2019}.\ 

To characterize cooling performance, we start by preparing a sample and detecting which tweezer contain an atom (see Sec.\,\ref{sec:overview}). 
We then cool for a time $t_{\rm cool}$ and perform release and recapture. Next, we cool the array before taking a final image to see which atoms survived and calculate the survival fraction.\ 

\begin{figure}
    \includegraphics[width=1\columnwidth]{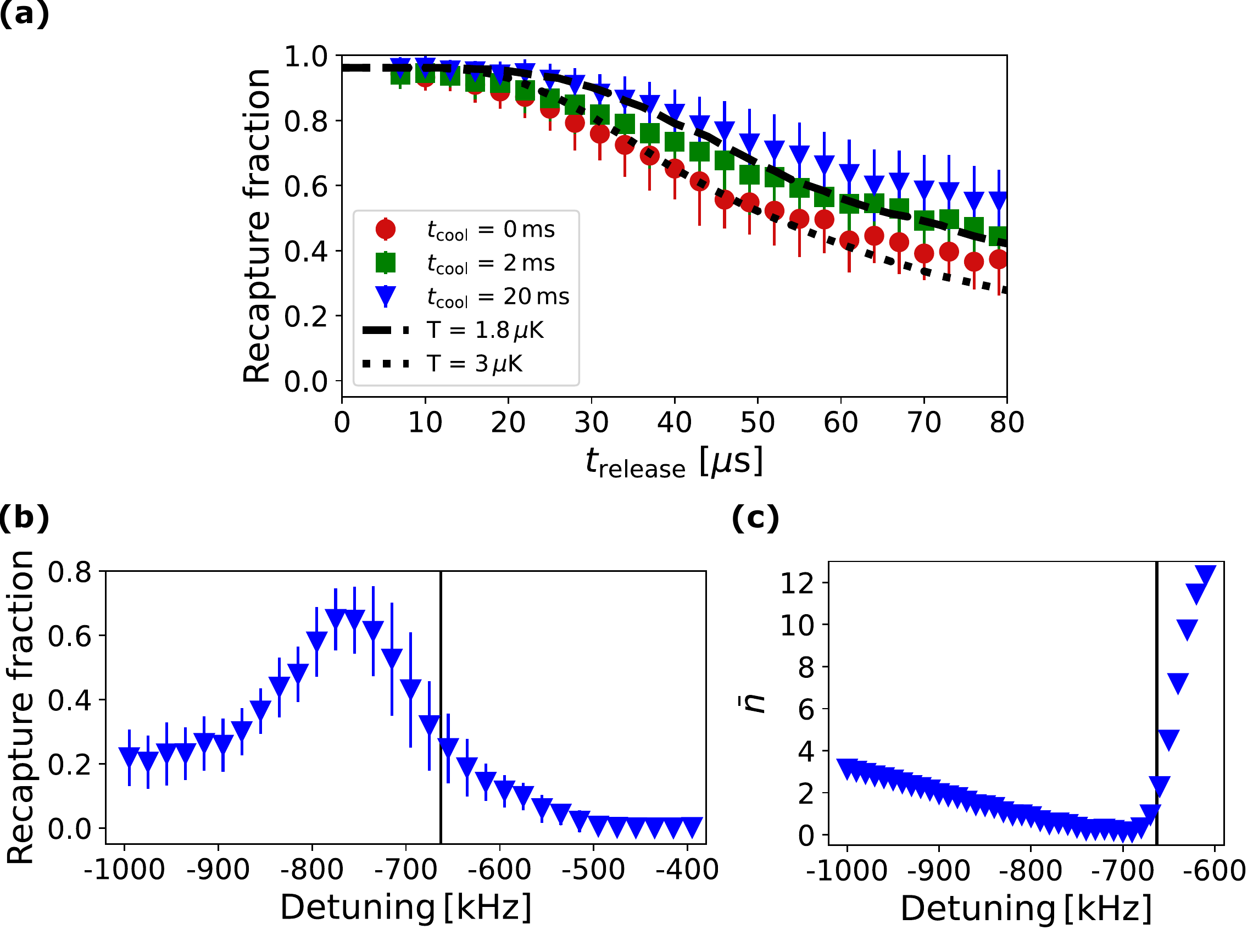} 
    \caption{\textbf{(a)} Temperature measurements using the release and recapture method for three different cooling times, $t_{\rm cool}$, after taking an image. The red circles show an atom directly after an image ($t_{\rm cool}$\,=\,0\,ms), green squares show a briefly cooled atom ($t_{\rm cool}$\,=\,2\,ms), and blue triangles show an optimally cooled atom ($t_{\rm cool}$\,=\,20\,ms). Error bars show the standard deviation calculated over the 36 atom array. All other errors fall well inside these error bars. The dashed (dotted) lines show the  results of Monte Carlo simulations for temperatures of 1.8\,$\mu$K (3\,$\mu$K) respectively.  \textbf{(b)} Recapture fraction of a single atom versus the cooling frequency for $t_{\rm cool}$\,=\,20\,ms and $t_{\rm release}$\,=\,60\,$\mu$s. As explained in Sec.\,\ref{sec:imaging}(c), brief cooling phases are interlaced in the imaging process, and here we vary the cooling frequency during imaging and during the period $t_{\rm cool}$. The vertical line shown at -663\,kHz indicates the approximate Stark shifted resonance of the cooling transition. \textbf{(c)} Average motional quanta $\bar{n}$, obtained by numerical simulation, in dependence of cooling light detuning at optimal intensity. The cooling transition is indicated as in (b).}
    \label{fig:CoolingParams}
\end{figure}

The results of such measurements under optimal cooling conditions for three $t_{\rm cool}$ are shown as examples in Fig.\,\ref{fig:CoolingParams}(a). The first measurement is taken directly after the first image ($t_{\rm cool}$\,=\,0\,ms, red circles). The second briefly cools the atoms ($t_{\rm cool}$\,=\,2\,ms, green squares). The third approaches the asymptotically coldest achievable temperature by using a long cooling time ($t_{\rm cool}$\,=\,20\,ms, blue triangles).\ 

Comparing this data with release and recapture simulations yields a temperature of approximately  1.8\,$\mu$K (dashed line in same figure)  for an optimally cooled atom ($t_{\rm cool}$=20\,ms), which is consistent with a temperature near the radial motional ground state energy of roughly $T=\frac{\hbar \omega}{2 k_b}\thicksim$1.1\,$\mu$K for our trap depth \cite{Tuchendler2008_release_recapture}. The release and recapture simulations are based on classical trajectories and could lead to an overestimation in the temperature as the atom approaches the motional ground state of the trap. Therefore we take this temperature estimate as an upper bound. \
 
To optimize cooling, we vary cooling light frequency or intensity while keeping $t_{\rm release}$\,=\,60\,$\mu$s and $t_{\rm cool}$\,=\,20\, fixed. Figure \,\ref{fig:CoolingParams}(b) shows an example of such a measurement for which the detuning is varied.
We find the highest recapture fraction, and therefore optimal cooling, for a frequency of $-775$\,kHz from the free space resonance and an intensity of $\thicksim\,88\,\,I_\text{sat}$ (Rabi frequency $\thicksim\,$2$\pi\times50$\,kHz).\ 

We compare the experimentally determined optimal parameters and the temperature with results of a numerical simulation of the cooling  process. 
The simulation is based on solving the steady state of a Lindblad master equation for a two-level atom in a pair of 1-D quantum harmonic oscillators (QHO), one for each  internal state $\ket{g}$, $\ket{e}$. 
The ratio of the  QHO frequencies is given by $\omega_g/\omega_e=\sqrt{\alpha_g/\alpha_e}=0.899$, with  $\alpha_{g,e}$  the dynamic polarizabilities at the tweezer wavelength. 
Choosing a traveling wave for the Sisyphus cooling laser, the transition dipole moments between  vibrational states of different QHO's are calculated as
$d_{eg}\bra{m}e^{ikx}\ket{n}$, with $\ket{m}$, $\ket{n}$ the vibrational states for internal states $\ket{g}$ and $\ket{e}$, respectively, and $d_{eg}$ the transition dipole moment of the \redtransition transition.\ 

We find the optimal parameters and the minimum temperature to be in good agreement (10$\%$) with the experimentally found ones. 
In Fig.\,\ref{fig:CoolingParams}(c) we show the number of average motional quanta after cooling in dependence of detuning at optimum intensity (all parameters of this simulation are given in Appendix C). 
At the minimum we obtain $\bar{n}\,\approx\,0.25$, which is in good agreement with measurements using sideband spectroscopy done by another group using the same cooling method \cite{Madjarov2020_Rydberg,Madjarov2019_clock,Choi2021_quantSim_endres}. \

\subsection{Optimizing the imaging parameters} 

We now discuss the imaging procedure and optimize its parameters. In a first approach we record the fluorescence of atoms while cooling. We find that the parameters that are optimal for cooling lead to a low scattering rate. The rate increases if the imaging beam frequency is chosen such that the atom is hotter. In the following we determine the imaging frequency and intensity that lead to highest scattering rate. We then explore a method to increase the fraction of atoms that survive imaging: interlacing imaging with brief cooling stages.\ 

The imaging frequency that leads to maximal scattering is found to be near the Stark shifted resonance (trap bottom). The scattering rate increases with imaging beam intensity approaching the theoretical maximum of $\thicksim23$\,kHz at our chosen operating intensity $I\thicksim\,350\,I_\text{sat}$. Deviations of $\pm50\%$ from this value have barely any effect on the scattering rate where lower intensities than this range cause a detectable decrease in the scattering rate away from the saturated regime. Higher intensities cause unnecessary heating of the atom and excess camera background noise during detection.
To clearly distinguish one atom from zero atoms, we image for 90\,ms (see Sec.\,\ref{sec:imaging}(D)).\

The scattering rate is maximized for different conditions than the ones leading to optimum cooling. The optimum cooling frequency is not close to the Stark shifted resonance, but approximately 2-3 radial motional sidebands to the red of the shifted resonance. This behavior is consistent with the fluorescence being suppressed by the Lamb-Dicke effect. The optimum cooling intensity ($\thicksim\,88 \,I_\text{sat}$) is much lower than the intensity used for imaging. Imaging is therefore accompanied by sub-optimal cooling, leading to a higher equilibrium temperature than optimum cooling, and potentially to higher atom loss.\ 

\begin{figure}
    \includegraphics[width=\columnwidth]{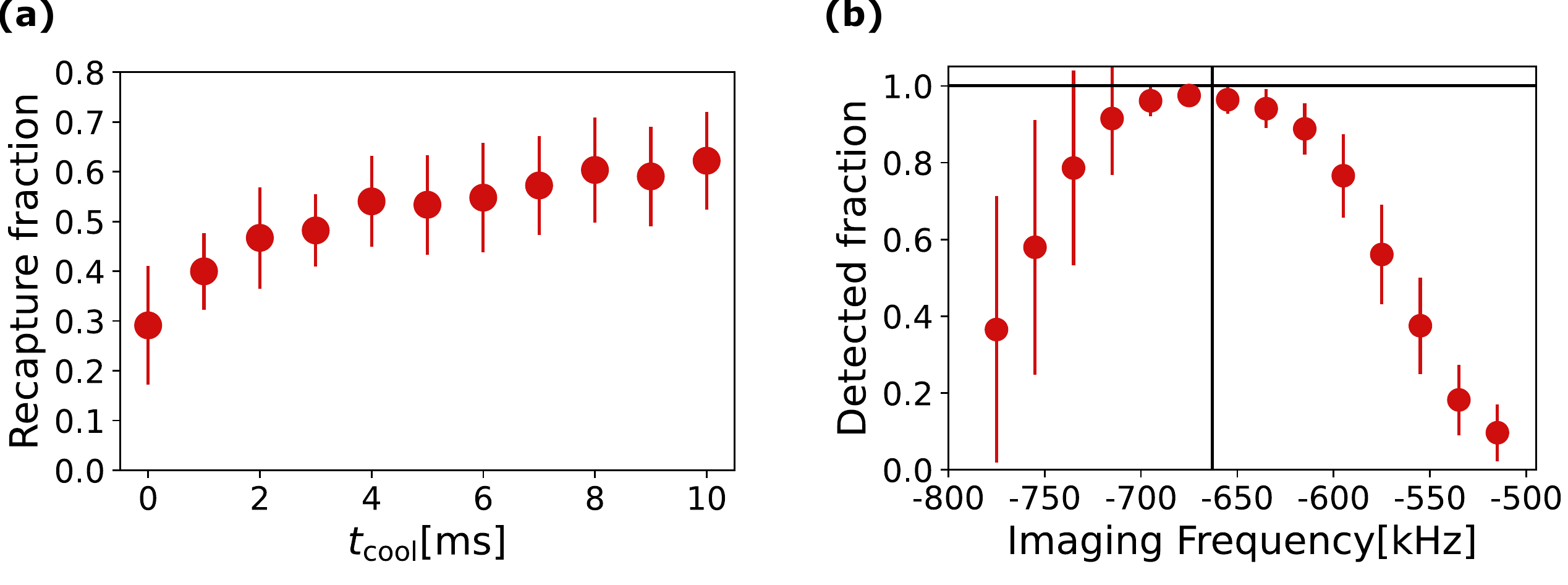}
    \caption{\textbf{(a)} Determination of cooling time scale. After heating the atoms for 90\,ms with imaging light only (detuning of -675\,kHz and $I\thicksim\,350\,I_\text{sat}$), we cool them for the time $t_{\rm cool}$ and then measure the survival fraction after release and recapture. We use a fixed release time of  $t_{\rm release}$\,=\,60\,$\mu$s. \textbf{(b)} Detected atom fraction versus imaging frequency. The cooling frequency is fixed at -775\,kHz from free space resonance of the \redtransition transition.  We use an optimized imaging duty cycle of 12$\%$ (88$\%$) cooling (imaging) light. The detuning is plotted with respect to  the free space resonance of the \redtransition transition.  The vertical solid line shows the approximate Stark shifted resonance of the transition. The point at -675\,kHz shows the highest survival probability of 0.97(2).}
    \label{fig:ImagingParams}
\end{figure}

We attempt to increase the fraction of atoms that survive imaging by interlacing imaging with cooling pulses. In order to determine how much cooling is needed we execute a single imaging pulse with the full duration needed for reliable single atom detection (90\,ms) followed by cooling. We estimate the temperature change during cooling by measuring the recapture fraction for a release time of 60\,$\mu$s, see Fig.\,\ref{fig:ImagingParams}(a). Cooling proceeds quickly for a few milliseconds, then approaches the steady-state for $t_{\rm cool}\gtrsim\,8$\,ms. This indicates that about 10$\%$ of the total imaging time should be spent on cooling to maintain a low temperature.\ 

To keep the temperature low during the imaging process, we interlace the 90\,ms of imaging time with eight cooling pulses of 1.5\,ms duration, i.e. we alternate eight times between 11\,ms of imaging and 1.5\,ms of cooling, each time changing frequency and intensity. 
This is the standard imaging timing sequence for all images in this work, unless stated otherwise. The duration of one cooling pulse was chosen to allow significant cooling while not wasting time at a low scattering rate for marginal additional cooling, see Fig.\,\ref{fig:ImagingParams}(a). The cooling pulse time is much longer than the timescales determining a single Sisyphus cooling cycle (axial and radial trap period, excited state lifetime, and inverse scattering rate) and allows the atom to scatter $\leq$\,34 photons.\

We now reoptimize the imaging frequency to maximize the fraction of detected atoms using imaging interlaced with cooling.  This fraction is measured by preparing a sample of single atoms using interlaced imaging with optimized operating parameters (see Sec.\,\ref{sec:overview}), and then determining how many atoms are also detected on a second interlaced image in dependence of the imaging frequency used for that image (see Fig.\,\ref{fig:ImagingParams}). 
For simplicity we use our standard threshold to distinguish zero and one atoms for all frequencies instead of optimizing it for every frequency. The best performance is reached for a detuning of -675 kHz from the free space resonance (-12\,kHz from the Stark shifted resonance), with a detected fraction of 0.97(2). We use this detuning for all images in this work unless stated otherwise. The benefit of imaging interlaced by cooling is that the detected fraction is $\thicksim$3$\%$ higher than what we could obtain without interlaced cooling. 

The reduction in the detected fraction for higher and lower detuning is due to different mechanisms. For blue detuning from the optimum value (right hand side of the plot in Fig.\,\ref{fig:ImagingParams}(b)) the reduction is dominated by the probability of an atom to be lost during imaging as evident from an increased variation in collected photon number. The blue detuned light can heat the atom out of the tweezer before sufficient photons can be scattered for detection. In the region red detuned of the optimum value  (left side of the plot) the poor detected fraction is instead dominated by insufficient scattering rate as evident from a decreased average number of collected photons. The reduction in scattering rate also leads to the drastically increased error bars at further red detuned frequencies. At these frequencies atoms do not scatter enough photons to clearly separate the single atom signal from the camera noise.\

We compare the temperature after our imaging process (which ends in a 1.5\,ms cooling stage) to the one obtained after a long cooling time (20\,ms additional cooling) using release and recapture measurements, see Fig.\,\ref{fig:CoolingParams}(a). Directly after the imaging process the data is well described by a simulation assuming 3\,$\mu$K (red circles and dotted line). This is not much above the temperature obtained after long cooling of about 1.8\,$\mu$K (blue triangles and dashed line).   \

\subsection{Detection fidelity, survival probability and minimum tweezer depth}

To distinguish tweezers containing one or zero atoms on fluorescence images we use a photon count detection threshold. We now illustrate this method and determine the optimum detection threshold and the detection fidelity. We measure the fraction of atoms that survive imaging and discuss its dependence on tweezer trap depth. Finally, we compare the lifetime of atoms under cooling and imaging conditions.\ 
\begin{figure}
    \includegraphics[width=0.8\columnwidth]{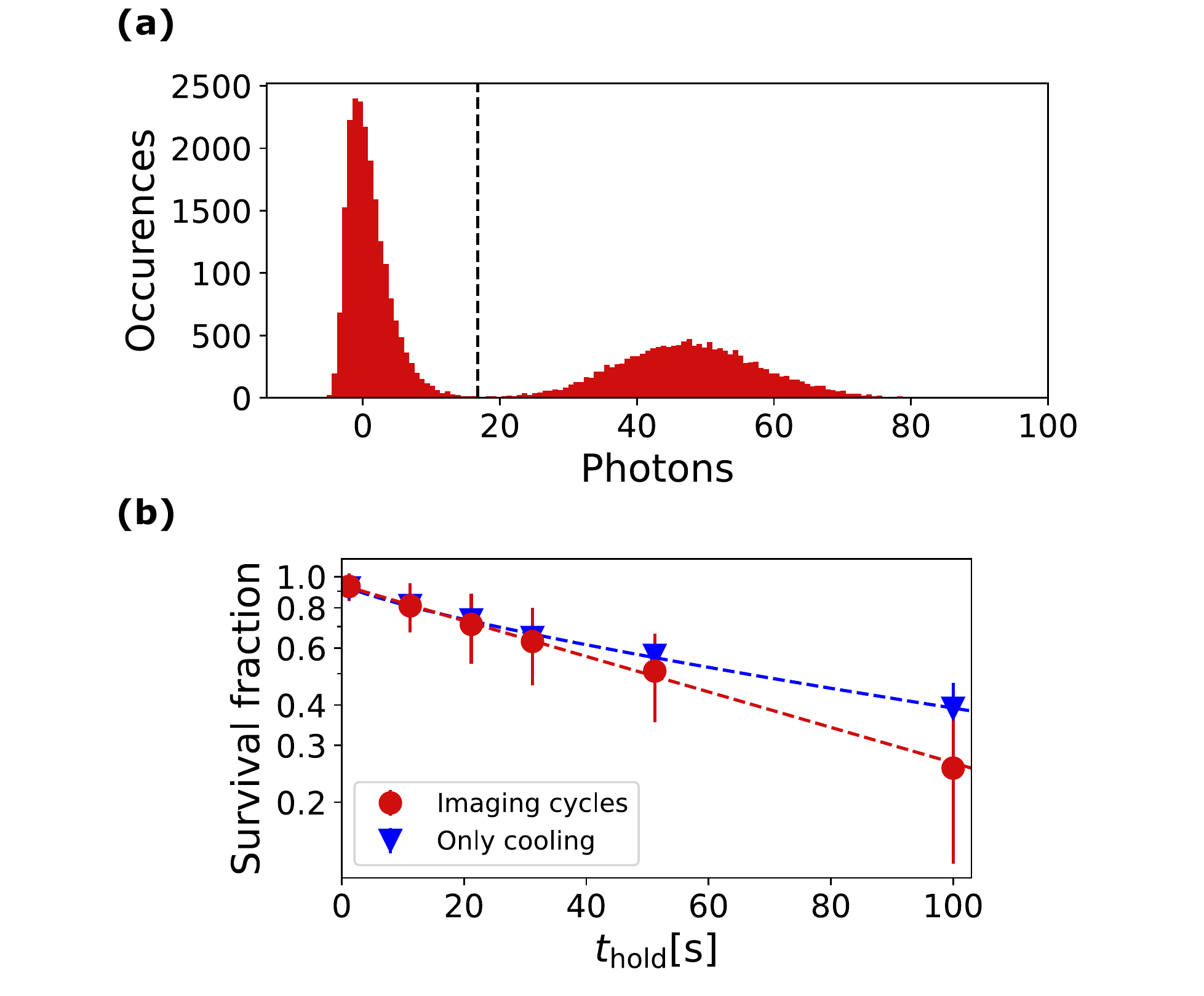}
    \caption{\textbf{(a)} A histogram showing  fluorescence photon counts from ROIs around single tweezers during imaging. This histogram combines the results of 1000 experimental runs using a 6$\times$6 tweezer array. The clear separation of the peaks highlights the uniform scattering over the array. The EMCCD counts per ROI have been converted to the number of incident photons. We use a bin size of 0.79 photons (150 EMCCD counts). The dashed line indicates the threshold separating the 0 atom peak (left) from the one atom peak (right) \textbf{(b)} The survival fraction versus the time spent  under optimized imaging cycles (red circles) and under optimal cooling (blue triangles). The dashed lines show fits to the data (see text).}
    \label{fig:HistoplusLifetime}
\end{figure}

%by $e^{-(t/\tau)^\alpha}$, where $\alpha$ and $\tau$ are fit parameters. The optimal cooling fit (green dashed line, $\alpha=0.8$) gives  $\tau$\,=\,116(5)\,seconds and the imaging fit (red dashed line, $\alpha=1$) gives $\tau$\,=\,79(3)\,seconds.
 
Figure \ref{fig:HistoplusLifetime}(a) shows a histogram of the number of collected photons in a tweezer ROI, where the average offset from background photons and camera noise is subtracted. Two distinct peaks are visible: one around zero photons, corresponding to no atom in the tweezer, and another around 50 photons, corresponding to the fluorescence count of a single atom. As is standard procedure \cite{cooper_alkaline-earth_2018,norcia_microscopic_2018, atom_computing_2021,Singh2021_dualspeciestweezers,Norcia_Clock_2019,covey_2000-times_2019,Madjarov2021_Thesis}, we postulate that an atom is present if the photon count is above a detection threshold, marked as dashed vertical line in the histogram. It may happen that randomly very few photons are scattered despite an atom being present in the tweezer or vice versa, leading to a wrong detection result. The detection fidelity is the probability of the detection to be correct. We determine it by calculating the overlap between a skewed Gaussian (fit to the zero atom peak) and a Gaussian (fit to the one atom peak) following the procedure outlined in \cite{Madjarov2021_Thesis}.  Using the detection threshold as an optimization parameter, we obtain a maximum detection fidelity of 0.9991(4) for a 135(20)\,$\mu$K trap depth.\

The duration of images can be decreased to 50\,ms with only a small loss in detection fidelity (fidelity reduced to 0.985). The loss is dominated by misidentifying a filled tweezer as empty at the optimal threshold, and is limited by background light on the camera and not by camera electronic noise. In particular the stray light of the repump lasers contributes to misidentification (the band pass filter in front of the camera insufficiently filters their light).  In fact, we obtain a better detection fidelity without using the repump lasers, also because optical pumping to metastable states happens rarely during imaging for our optical tweezer intensities. This is the reason why we do not use the repump lasers during our imaging process for shallow traps.\ 

To determine the probability of an atom to survive the imaging process we record two images in sequence. The probability to detect an atom on the second image if it was present on the first is 0.97(2). The trap depth can be reduced to 99(15)\,$\mu$K without sacrificing detection fidelity. However, once the tweezer trap depth is decreased below the $\thicksim$\,135\,$\mu$K level, the chance of recovering the atom on a second image starts to decrease. For example we measure a decrease in survival probability to 0.926(65) for a trap depth of 99(15)\,$\mu$K.\ 

Survival probability is reduced for trap depths below $\thicksim$\,135\,$\mu$K because one of the Sisyphus cooling criteria outlined in Sec.\,\ref{sec:imaging}(A) is not met. For such low trap depths, the excited state $axial$ trap frequency becomes lower than the natural linewidth of the transition ($\thicksim\,7.4$\,kHz). When the axial trap frequency becomes that small the cooling process does not sufficiently compensate fluorescence recoil heating in the axial direction. The fact that Sisyphus cooling works for higher trap depths highlights the ability of the single radial cooling beam to remove energy from all directions simultaneously. Imaging at even lower trap depths could be achieved by using closer to spherically symmetric  potentials as those in \cite{Young2020_halfminuteClock} or in a 3D lattice \cite{Schine2021_Lattice}, allowing reliable imaging at even lower trap depths. Already the achieved trap depth of 135\,$\mu$K for reliable imaging is three times less than obtained with blue imaging \cite{covey_2000-times_2019}, making it possible for us to obtain three times more tweezers for a given tweezer laser source power.\ 

In Figure \ref{fig:HistoplusLifetime}(b) we show the survival probability over time $t_{\rm hold}$ when continually imaging or when just cooling. For these long measurements we turn on the repump lasers and close the atomic beam shutter. 
We fit the data by $e^{-(t/\tau)^\alpha}$, where $\alpha$ and $\tau$ are fit parameters.
For continual imaging cycles the fit provides $\alpha$=1 and a 1/e lifetime of $\tau$=79(3) seconds, allowing for hundreds of pictures to be taken of a single atom. This decay of the survival fraction is equivalent to $p_1^N$, where $N$ is the number of elapsed images and $p_1=0.9986(4)$ \cite{covey_2000-times_2019}. For continuous cooling the fit provides $\alpha$=0.8 and $\tau$=116(5)\,seconds. The deviation from a pure exponential decay might be due to slowly improving vacuum quality over the course of each measurement, triggered by the atomic beam shutter closure. When analysing individual tweezers, we find that some tweezers have the same lifetime under cooling and imaging conditions.\ 

The finite lifetime can have a variety of origins. We verify that the temperature of the atoms stays constant under both investigated conditions, excluding a slow process heating the atoms out of the trap. We find that the lifetime depends on the vacuum quality, as lifetime degrades over months and increases to the values stated above only after flashing titanium sublimation pumps. The decrease in lifetime from cooling to imaging conditions for most tweezer sites indicates that the small trap depth variation between tweezers of $3\%$ make it impossible to optimize cooling and imaging for all tweezers.\ 

We observe day to day changes of the survival probability originating from drifts away from ideal conditions, in particular magnetic field drifts. Magnetic field drifts on the $\thicksim$\,20\,mG level affect the single image survival probability significantly ($\thicksim\,2$\% reduction).

\section{Site selective imaging}
\label{sec:siteselective}
 
Our imaging technique provides an additional advantage. 
Using an easily achievable differential Stark shift, one can tune a certain tweezer out of resonance with the imaging light used for the rest of the array. This allows for selective imaging of either the remaining  tweezers of the array or of the single shifted tweezer. The ability to selectively readout a  single atom from the array is a necessary step for error correction in many quantum computation algorithms \cite{Anderson2021_ionQEC,Sohn2019_QEC}.\ 

\begin{figure}[t]
        \includegraphics[width=0.95\columnwidth]{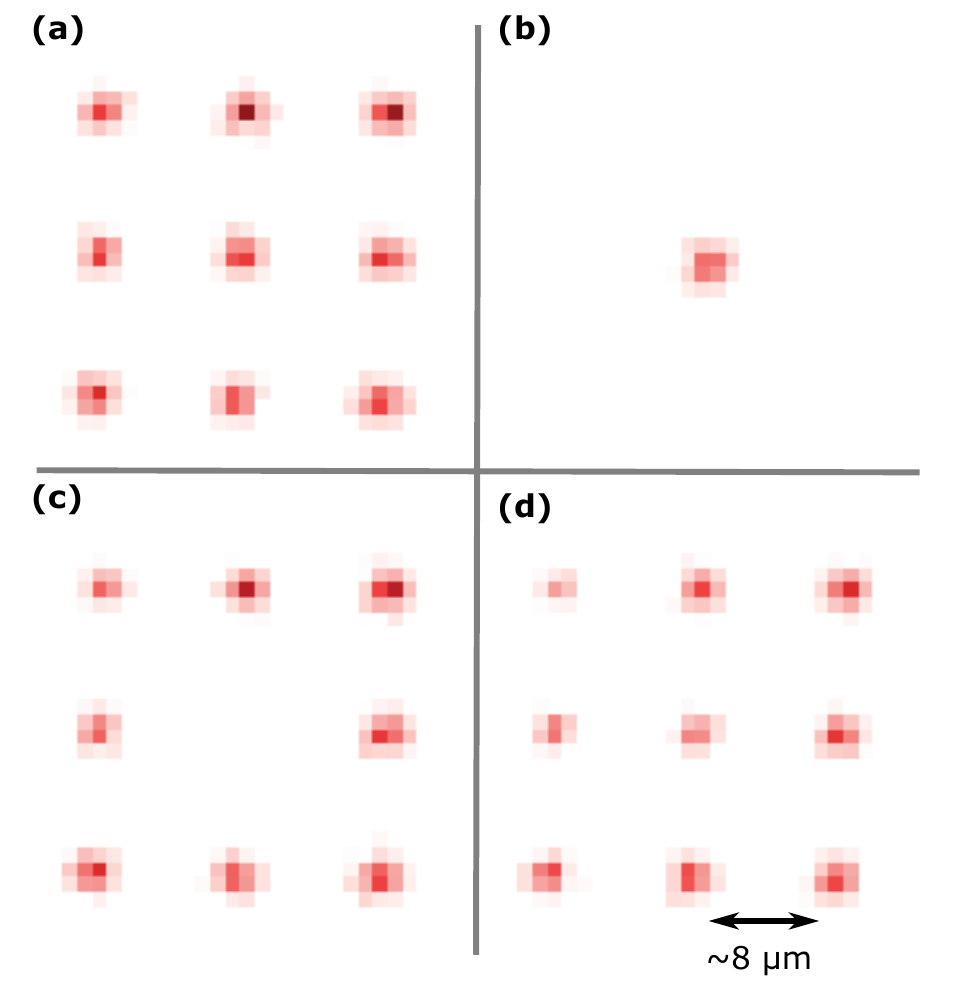}
    \caption{Site selective imaging using an additional tweezer to apply a differential Stark shift to one tweezer. \textbf{(a)} A first image is recorded to check loading. \textbf{(b-c)} After turning on the AOD tweezer on top of the center tweezer, we image first only the center tweezer and then the rest of the array by using respectively appropriate imaging and cooling detunings. \textbf{(d)} When the AODs are turned off, all traps are visible in a final image using the normal imaging detuning. The images show the average fluorescence from 300 experimental realizations.}
    \label{fig:darkout}
\end{figure}

To demonstrate  the ability to select (or dark out) an atom from the image, we use a tweezer created by the crossed AODs to create a deeper potential for a single tweezer site in a 3$\times$3 tweezer array.  
To characterize site selective images, we record four consecutive images in one experimental run, see Fig.\,\ref{fig:darkout}. The first image (Fig.\,\ref{fig:darkout}(a)) is taken directly after single atom preparation, as described in Sec.\,\ref{sec:overview}.
The AODs are then turned on for the second and third image (Fig.\,\ref{fig:darkout}(b,c)), in which we image the single shifted tweezer and the rest of the array respectively. 
To record atoms in the shifted tweezer, we increase the imaging detuning to -2\,MHz (i.e. 1.325\,MHz to the red of the usual imaging detuning).
In the fourth image we turn off the additional tweezer and again image the entire array (Fig.\,\ref{fig:darkout}(d)).\ 

We can image the single site such that it is detected in image two with a survival probability  of  0.96(2), and never appears in image three. Moreover,  the entire atom array survives this 'dark out' measurement with a probability of 0.95(2). 

 The lower survival probability, in comparison to the value obtained in Sec.\,\ref{sec:imaging}, is due mainly to worse balancing of the trap intensities for the nine trap array used for this measurement. Additionally, the cooling frequency and intensity in the deeper single tweezer were not fully optimized.
 We note that the AC Stark shift chosen here is too small to fully protect quantum information of the remaining atoms but provides an initial proof of concept. However a stronger AC Stark shift could make this feasible.

\section{Conclusion}
\label{sec:conclusion}
In conclusion, we have demonstrated the detection of single Sr atoms in shallow tweezers with high fidelity (0.9991(4)) and survival probability (0.97(2)). Detection is based on imaging on the red, narrow linewidth \redtransition\,$(|m_j|=1)$\ transition. We show that with proper frequency and intensity modulation a high scattering rate can be maintained while keeping the temperature of the atoms low.
 
Our red imaging technique works for a wide range of trap depths, and for shallow traps, red imaging is advantageous over blue imaging. We need slightly ($\thicksim$2-fold) increased imaging times (100\,ms instead of 50\,ms) in comparison to blue imaging on the broad linewidth \bluetransition transition in deep traps (450\,$\mu$K depth) \cite{covey_2000-times_2019}. However, in shallow traps blue imaging is limited by the cooling rate, leading to excessive imaging times in comparison to red imaging \cite{Schine2021_Lattice}.

In contrast to blue imaging, red imaging avoids optical pumping of ground state atoms into the metastable states (via $^{1\hspace{-0.3ex}}D_2$). Imaging in shallow traps reduces off-resonant scattering of trap light by metastable state atoms ($^{3\hspace{-0.3ex}}P_{0,2}$), leading to longer coherence times. Red imaging in shallow traps combines both advantages and enables high-fidelity shelving into metastable states for state specific detection or clock readout.

We show that, with a small additional Stark shift, we can isolate a single tweezer of the array from the imaging process. This allows us to selectively image (or hide) a single atom of the array.
Through application of a bias field of $\thicksim$\,50\,G, this selective imaging technique could be further extended to \textit{state-selective} imaging for hyperfine ground states in the fermionic isotope. 
This opens the possibility of  imaging more than two hyperfine ground states without disturbing the others. This will be a useful tool for quantum simulations or qudit style quantum computing \cite{Weggemans2021_Qudit,Omanakuttan2021_qudit,Banerjee2013_Su(N)}.

It should be possible to extend red imaging to situations beyond the specific one examined here. The small potential wells containing the atoms can also be created by optical lattices, or other tightly confining dipole traps, making it possible to use the technique in quantum gas microscopes or 3D lattice clocks. 
This detection technique should work at nearly all tweezer wavelengths where one of the $m_J$ states of $^{3\hspace{-0.3ex}}P_1$ is stronger trapped than the ground state. In particular, using 515-nm tweezers would be an appealing option. This is because red imaging avoids the leakage channel through $^{1\hspace{-0.3ex}}D_2$ from which blue imaging suffers \cite{cooper_alkaline-earth_2018,norcia_microscopic_2018}. Tweezers at this wavelength are also likely to trap  most Rydberg states \cite{Madjarov2021_Thesis}. The large polarizabilities at this wavelength, small diffraction limit of the tweezer light, and shallow required trap depth of red imaging would allow for the creation of $\ge\,1000$ strontium atom tweezer arrays with current laser technology.

\section*{Acknowledgments}
We  thank the other members of the Strontium Quantum Gasses group at the University of Amsterdam for useful discussions and debugging tips. We also thank the Eindhoven quantum computing group led by Servaas Kokkelmans and Edgar Vredenbregt for stimulating discussions.
This project is supported by the Netherlands Organization for Scientific Research (NWO) under the Gravitation grant No. 024.003.037, Quantum Software Consortium, and under grant No. FOM-153.
This work is supported by the Dutch Ministry of Economic Affairs and Climate Policy (EZK), as part of the Quantum Delta NL programme. \\

\section*{Appendix A: Experimental Sequence}

\begin{figure}[t]
    \includegraphics[width=\columnwidth]{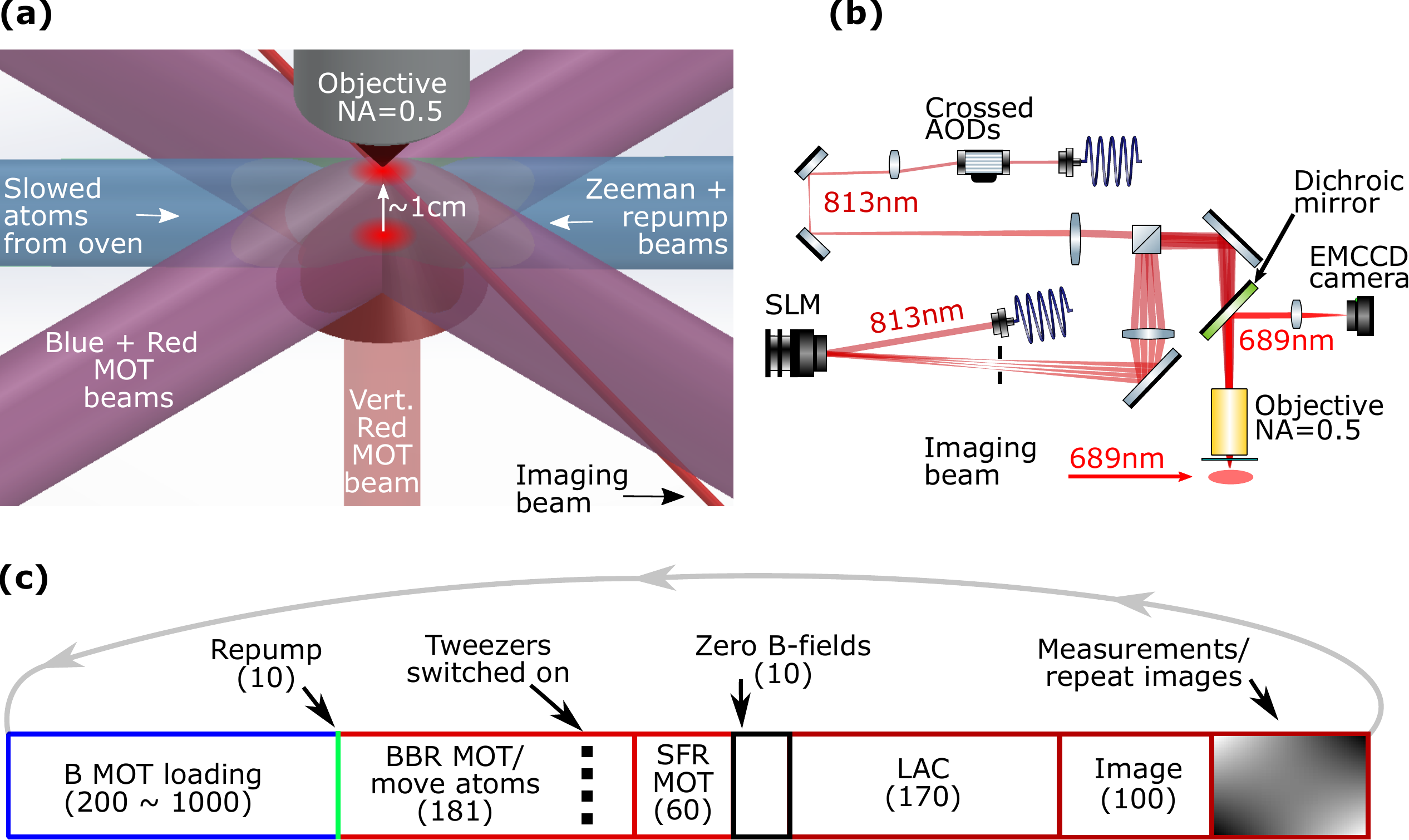}
    \caption{\textbf{(a)} Detailed sketch of crucial elements of the experimental setup. Thite arrow from the left indicates the beam of slowed atoms from the oven. The Zeeman slower and repump beams are shown in blue. The horizontal, overlapping blue and red MOT beams are shown in purple. An additional vertically upwards propagating red MOT beam (shown in pale red) provides confinement against  gravity. The initial and final red MOT positions are shown as intense red spots. The red MOT is moved vertically 1\,cm into the focus of the microscope objective. The 679 nm and 707 nm repump lasers co-propagate with the cooling/imaging beam.    \textbf{(b)} Schematic of the  tweezer setup. The main tweezer array is generated using an SLM. An additional tweezer can be created using a pair of crossed AODs. The SLM and AOD tweezers are combined using a polarizing beam splitter. Both systems are imaged through an NA=\,0.5 objective onto the atoms. Fluorescence light from the atoms is separated from the tweezer light using a long pass dichroic mirror. The fluorescence is then sent onto an EMCCD camera (Andor Ixon 897).      \textbf{(c)} Experimental sequence (numbers in brackets give time spans in ms). The figure uses acronyms for blue MOT (B MOT), broadband red MOT (BBR MOT), single frequency red MOT (SFR MOT)  and light assisted collisions (LAC).}
    \label{fig:SuppInfo}
\end{figure}

We utilize a unique technique for loading our narrow linewidth MOT in order to create optical access for the microscope objective. First, $^{88}$Sr atoms from an  $\thicksim$\,$500^{\circ}\mathrm{C}$ oven are slowed using a Zeeman slower operating on the 2$\pi\times30$ MHz wide \bluetransition transition at 461\,nm. 
The slowed atoms are then  further cooled and compressed by a four beam 'blue' MOT, also using the \bluetransition transition,  in a 3D quadrupole field to milli-Kelvin temperatures.
This blue MOT consists of two sets of retro-reflected beams ($1/e^2$ waist of $\thicksim$\,12\,mm) that are perpendicular to each other and horizontal.
Refraining from implementing the usual third MOT beam pair allows us to place the microscope objective along the gravity axis without complications from that beam pair, as shown in Fig.\,\ref{fig:SuppInfo}(a). 
This blue MOT is an incomplete trap as it provides no confinement against gravity.
However the blue MOT is able to quickly cool and confine the atoms in the horizontal plane, which comprises the only dimension along which atoms entering from the Zeeman slower are fast. 
A significant fraction of atoms are then trapped in the quadrupole magnetic field (52\,G/cm gradient in the axial direction, which is vertical) of the MOT by optical pumping to low-field seeking states of the $^{3\hspace{-0.3ex}}P_2$ manifold.
This optical pumping is naturally happening when atoms rapidly scatter MOT light and decay from $^{1\hspace{-0.3ex}}P_1$ through the 5s4d $^{1\hspace{-0.3ex}}D_2$ state to $^{3\hspace{-0.3ex}}P_2$. 

After quadrupole trap loading, all blue lasers are switched off and the magnetically trapped atoms are repumped back to the ground state, using a 497\,nm laser resonant with the $^{3\hspace{-0.3ex}}P_2$-$^{3\hspace{-0.3ex}}D_2$ transition. Simultaneously the quadrupole field gradient is reduced to 0.63\,G/cm in the vertical direction. The atoms are then loaded into a five beam narrow linewidth "red" MOT operating on the \redtransition \ transition. 
Four of the five red MOT beams are overlapped with the blue MOT beams and the fifth beam ($1/e^2$ waist of $\thicksim$6\,mm) is propagating vertically upwards.

There is no need for a downwards propagating beam because the upward radiation pressure force is limited by the narrow linewidth and the MOT quadrupole field to a small phase-space region. This force is counter-balanced by gravity. The atoms settle into a cloud on the lower part of a shell of equal B-field magnitude below the quadrupole centre. This shell is defined by the detuning of the MOT beams being equal to the Zeeman shift induced by the B-field. This trap scheme again does not need a beam going through the microscope
objective.\ 

The red MOT beams are initially frequency modulated in order to create a comb of frequencies from $\text{-60}$ to $\text{-3000}$ kHz detuning with 20 kHz spacing.
%This  increases the capture efficiency of the initial red MOT by better matching the spatial and temporal distribution of the atoms in the magnetic reservoir.  
The modulation range and intensity of this broadband red MOT are decreased over 181 ms, while  a bias field of $\thicksim$0.6\, G against gravity is ramped on, raising the atoms by 1\,cm, from the centre of the vacuum chamber to the focal plane of the microscope objective, by shifting the center of the quadrupole field. An additional small bias field produced by three orthogonal coil pairs is ramped while the MOT position is raised and used to finely position the red MOT onto the tweezer array. The frequency modulation  is then switched off and single frequency red MOT beams, with a detuning of $-100$\,kHz  and intensity of 8\,$I_{\rm sat}$, are used to load the tweezers. \

We optimize all parameters of the experiment up to this point on achieving the desired red MOT atom number in a reliable way and in a short time.
We find that for MOTs (with our selected detuning) of 5$\times10^4$ to 3$\times10^6$ atoms, the entire tweezer array can be loaded  with $\geq$\,1 atom per site, where on the low end we get slightly below unity filling.
On the high end the high density of atoms in the tweezers leads to less than half of the tweezers being filled with atoms, presumably because of additional \textit{non}-pairwise losses during the light assisted collision step.
Because of this robustness to atom number fluctuations, we load the magnetic reservoir for a variable amount of time at the end of  an experimental sequence ($\thicksim\,200$\,ms). 
The exact time is determined by the need to store data of the last run on the data analysis computer and the need to prepare the next experimental sequence in the experiment control computer.
This procedure creates a MOT of approximately 5$\times10^5$  atoms at a temperature $\leq\,1.5\,\mu$K in the focal plane of our objective.\

Slightly before the MOT is switched to single frequency operation, the tweezers are switched on, see also next Appendix.
After the red MOT has reached its final position we wait for 50\,ms to load the tweezers. Then we switch off the MOT lasers and quadrupole field,  and  ramp the bias magnetic fields to 0\,G at the position of the tweezers.\ 

Single atom detection only  worked properly once we
spectrally filtered  the 689-nm laser light used to induce fluorescence. 
The source of all 689-nm light is an external cavity diode laser (ECDL) that is short-term stabilized on a reference cavity with a linewidth of 35\,kHz, which in turn is long-term stabilized (in length) on a spectroscopy signal. 
Light from the ECDL is amplified by injection locked lasers and then used on the experiment. 
Initially we used light from the ECDL directly to inject the  amplifying diodes. 
This ECDL light is spectrally broadened by servo bumps from the locking electronics and amplified spontaneous emission and we found it impossible to prepare and detect single atoms. 
We then used the light that is transmitted through and therefore filtered by the reference cavity to inject the amplifying diodes, allowing us to achieve the single atom preparation and detection results presented here. 
The red MOT behavior did not noticeably change when switching from unfiltered to filtered light.\

The light used for light assisted collisions, imaging, and cooling  is sent onto the tweezer array via a single beam with a polarization perpendicular to both the tweezer
propagation axis and the tweezer polarization, and a $1/e^2$ waist of $\thicksim$\,1\,mm. 
We find robust single atom preparation of $\thicksim$\,50\% for a detuning of approximately -100\,kHz from the Stark shifted resonance (-750\,kHz from free space resonance). We change the intensity of the beam from low ($I\,\thicksim\,88\,I_\text{sat}$ for 10\,ms) to high ($I\,\thicksim\,700\,I_\text{sat}$ for 150-200\,ms) then back to low ($I\,\thicksim\,88\,I_\text{sat}$ for 10\,ms) in order to cool the loaded atoms into the tweezer, induce light assisted collisions \cite{cooper_alkaline-earth_2018,Schlosser2002_CollisionalBlockade}, and cool the single atom before taking the first image of an experimental run.

\section*{Appendix B: tweezer creation}
The 813.4-nm laser light used for the optical tweezers in our experiment is generated by an external cavity diode laser, which is amplified to 1.7 W using a tapered amplifier (TA). The output of the TA is divided into two optical paths, a main path to create the tweezer array using an SLM (Meadowlark P1920 1920$\times$1152) and a second path for a movable tweezer using AODs (AA opto-electronic DTSXY-400-800), see Fig.\,\ref{fig:SuppInfo}(b). The main output path is sent through a dispersive prism in order to filter out any amplified spontaneous emission from the TA and is then sent through an acousto-optic modulator for intensity  control before being coupled into a fibre. The second path is sent without further filtering into an optical fibre. \ 

The optical tweezers are created by imaging an array of beams through a microscope objective (NA=\,0.5, Mitutoyo 378-848-3). An almost arbitrary and stationary pattern of tweezers is created using the SLM. In order to calculate the phase pattern of the desired tweezer pattern we use the weighted Gerchberg-Saxton algorithm  \cite{Nogrette2014}.
The phase imprinted onto the incident beam by the SLM is a sum of phases including the tweezer array pattern phase, a lens phase to Fourier transform the phase to a real image,  a grating phase to separate the zeroth order,  and a factory correction phase. \ 

The sum of these phases creates an array of foci $\thicksim$\,180\,cm from the SLM.
This array is imaged through the microscope objective (effective focal length $f=4$\,mm) with a field lens of $f=500$\,mm taking care that the array of beams is conjugated onto the aperture of the microscope objective.
In this work all results shown have been performed with the SLM creating a square 6$\times$6 array of tweezers unless otherwise noted.\

Additional balancing of the tweezer trap depths can  be achieved by finetuning the SLM pattern. As a first step we spectroscopically measure the depth of each tweezer by inducing heating on the \redtransition\,$(m_j = 0)$ transition, which is weaker trapped than the ground state. The loss feature is then fit with a Gaussian, and the center frequency is extracted for each tweezer. The detuning of this frequency from the free space resonance is proportional to the tweezer intensities. The amplitude of each tweezer, in the pattern to be calculated, is then weighted based on these measured center frequencies. The tweezer phase pattern is then recalculated using these new weights.  This procedure allows us to balance the trap depths across the 6$\times$6 array to a standard deviation of approximately 3\% \cite{Tuchendler2008_release_recapture,Singh2021_dualspeciestweezers,cooper_alkaline-earth_2018}. This procedure was not executed for the 3$\times$3 array used in Sec.\,\ref{sec:siteselective}. \

We finally characterize the depth of our tweezers using spectroscopic method explained above. The error in our trap depth determination has two sources. The choice of either the blue edge frequency or center frequency, based on if a purely thermally broadened line shape or a purely power broadened line shape is fit respectively, provides 15\,$\mu$K of error \cite{cooper_alkaline-earth_2018}. An additional 5\,$\mu$K uncertainty comes from the 3\,$\%$ standard deviation of the optimized SLM pattern.
For the tweezers used throughout the paper, we estimate a waist of $\thicksim$\,0.84\,$\mu$m and an optical power on the atoms of $\thicksim$\,2.33\,mW per tweezer.\ 

An additional tweezer (or tweezers) can be created using a crossed pair of acousto-optic deflectors (AODs) whose position in the focal plane  can be controlled through radio frequency tones sent to the AODs. 
Unlike the SLM's slow refresh rate, this tweezer can move at speeds sufficient for sorting atoms into defect free arrays, or for quickly applying an additional tweezer for isolating one tweezer from the rest of the array (see Sec.\,\ref{sec:siteselective}).\\

\section*{Appendix C: Sisyphus cooling simulation parameters}
For the Sisyphus cooling simulation we consider  a trap depth of $k_B \times$135\,$\mu$K = $h \times 2.8$\,MHz. 
For our estimated waist of 0.84\,$\mu$m, this gives a \textit{radial} trap frequency of $43$\,kHz for the ground state. 
We calculate the $^1S_0$ $(\alpha_g)$ and $^{3\hspace{-0.3ex}}P_1\,(|m_j|=1,\alpha_e)$ polarizabilities to be 286\,a.u. and 355\,a.u. respectively. 
For the results presented in Fig.\,\ref{fig:CoolingParams}(c), we use a Rabi frequency of 2$\pi\times42$\,kHz. We include 15 harmonic oscillator levels in our calculation.     
\bibliography{redimg}

\end{document}